\documentclass{PoS}

\usepackage{graphicx}

\newcommand{\half}{{\frac{1}{2}}}
\newcommand{\mbf}[1]{\mathbf{#1}}
\renewcommand{\bar}[1]{\overline{#1}}

\title{Gauge/Gravity Duality and Strongly Coupled Light-Front Dynamics}

\ShortTitle{Gauge/Gravity Duality and Strongly Coupled Light-Front Dynamics}

\author{\speaker{Guy F. de T\'eramond}
 \\
       Universidad de Costa Rica, San Jos\'e, Costa Rica\\
       E-mail: \email{gdt@asterix.crnet.cr}}

\author{Stanley J. Brodsky\\
 SLAC National Accelerator Laboratory\\
Stanford University, Stanford, CA 94309, USA, and \\
CP$^3$-Origins,
Southern Denmark University,
Odense, Denmark\\
 E-mail: \email{ sjbth@slac.stanford.edu}}

\abstract{We find a correspondence between semiclassical gauge theories quantized on the light-front and a dual gravity model in anti-de Sitter (AdS) space, thus providing an initial approximation to QCD in its strongly coupled regime. This correspondence -- light-front holography -- leads to a light-front Hamiltonian and relativistic bound-state wave equations in terms of an invariant impact variable $\zeta$ which measures the separation of the quark and gluonic constituents within the hadron at equal light-front time.
Light-front holography also allows a precise mapping of transition amplitudes from AdS to physical space-time. In contrast with the usual AdS/QCD framework, the internal structure of hadrons is explicitly introduced in the gauge/gravity correspondence and the angular momentum of the constituents plays a key role. We also discuss how to introduce higher Fock-states in the correspondence as well as their relevance for describing the detailed structure of space and time-like form factors.}

\FullConference{Light Cone 2010: Relativistic Hadronic and Particle Physics\\
		 June 14-18, 2010\\
		 Valencia, Spain}

\begin{document}

\section{Introduction}

The AdS/CFT correspondence~\cite{Maldacena:1997re} between
string states on Anti--de Sitter (AdS) space-time and conformal  field theories (CFT) in physical space-time has
brought a new perspective for the study of the dynamics of strongly coupled quantum field theories and has led
to new analytical insights into the confining dynamics of QCD, which is difficult to realize using other methods.
In practice, the duality provides an effective gravity description in a ($d+1$)-dimensional AdS
space-time in terms of a flat $d$-dimensional conformally-invariant quantum field theory defined at the AdS
asymptotic boundary.~\cite{Gubser:1998bc} Thus, in principle, one can compute physical observables
in a strongly coupled gauge theory  in terms of a classical gravity theory.

The original correspondence~\cite{Maldacena:1997re}  is the duality between
${\cal N}=4$ supersymmetric SU$(N_C)$ Yang-Mills theory (SYM) and
the supergravity approximation to Type IIB string theory on  AdS$_5 \times S^5$ space.
QCD is fundamentally different from SYM
theories where all the matter fields transform in adjoint multiplets of SU$(N_C)$. Unlike SYM theories, where the conformal invariance
implies that the coupling does not run with energy,   its scale invariance is broken in QCD by quantum effects.

The most general group of transformations that leave the AdS metric
\begin{equation} \label{AdSz}
ds^2 = \frac{R^2}{z^2} \left(\eta_{\mu \nu} dx^\mu dx^\nu - dz^2\right),
\end{equation}
invariant, the isometry group, has dimensions $(d+1)(d+2)/2$. Thus, for $d=4$, five dimensional anti-de Sitter space AdS$_5$ has 15 isometries, in agreement with the number of generators of the conformal group in four dimensions. The metric (\ref{AdSz}) is invariant under the transformation $x \to \lambda x$, $z \to \lambda z$. The variable $z$ is thus like a scaling variable in Minkowski space: different values of $z$ correspond to different energy scales at which the hadron is examined.

A gravity dual of QCD is not known, and it has proven difficult to extend the gauge/gravity duality beyond theories which are to a great extent constrained
by their symmetries. We shall follow here a simplified approach, which is  limited to the study of the propagation of hadronic modes in a fixed effective gravitational background
 which encodes salient properties of the QCD dual theory, such
as the ultraviolet conformal limit at the AdS boundary at $z \to 0$, as well as modifications of the background geometry in the
large $z$ infrared region from confinement.   The introduction of an infrared  cutoff at a finite value   $z_0 \sim \Lambda_{\rm QCD}$ is a simple way to get confinement and  discrete
normalizable modes.  Thus the ``hard-wall'' at $z_0$ breaks conformal invariance and allows the introduction of the QCD scale  and a spectrum of particle states.~\cite{Polchinski:2001tt}  As first shown by Polchinski and Strassler,~\cite{Polchinski:2001tt} the AdS/CFT duality, modified
to incorporate a mass scale, provides a derivation of dimensional counting
rules~\cite{Brodsky:1973kr,  Matveev:ra} for the leading
power-law fall-off of hard scattering beyond the perturbative regime.
The modified theory generates the hard behavior expected from QCD, instead of the soft
behavior characteristic of strings.

In the usual AdS/QCD approach~\cite{Erlich:2005qh, DaRold:2005zs} bulk fields are introduced to match the
$SU(2)_L \times SU(2)_R$ chiral symmetries of QCD and its spontaneous breaking, but without explicit connection with the internal constituent structure of
 hadrons.~\cite{Brodsky:2003px}  Instead, axial and vector currents become the
primary entities as in effective chiral theory.
The conformal metric of AdS space can be  modified  within the gauge/gravity framework with the introduction of a dilaton field to reproduce the observed linear Regge behavior in the hadronic spectrum.~\cite{Karch:2006pv} The additional warp factor in the metric, or, equivalently, the introduction of a dilaton
background $\varphi(z)$  introduces an energy scale in the five-dimensional Lagrangian, thus breaking the conformal invariance.
A particularly interesting case is a dilaton profile $\exp{\left(\pm \kappa^2 z^2\right)}$ of either sign, the ``soft-wall'', since it
leads to linear Regge trajectories~\cite{Karch:2006pv} and avoids the ambiguities in the choice of boundary conditions at the infrared wall.

Light-front quantization is the ideal framework to describe the
structure of hadrons in terms of their quark and gluon degrees of
freedom. The simple structure of the light-front (LF) vacuum allows an unambiguous
definition of the partonic content of a hadron in QCD and of hadronic light-front wavefunctions (LFWFs)
which relate its quark
and gluon degrees of freedom to their asymptotic hadronic state. The LFWFs of relativistic bound states in QCD provide a description of the structure and internal dynamics of hadronic states in terms of their constituent quark and gluons  at the same LF time  $\tau = x^0 + x^3$, the time marked by the
front of a light wave,~\cite{Dirac:1949cp} instead of the ordinary instant time $t = x^0$. The constituent spin and orbital angular momentum properties of the hadrons are also encoded in the LFWFs. Unlike instant time quantization, the Hamiltonian equation of motion in the light-front is frame independent and has a structure similar to the eigenmode equations in AdS space.
This makes a direct connection of QCD with AdS/CFT methods possible. The identification of orbital angular momentum of the constituents is a key element in our description of the internal structure of hadrons using holographic principles,
since hadrons with the same quark content but different orbital angular momentum have different masses.

A physical hadron in four-dimensional Minkowski space has four-momentum $P_\mu$ and invariant
hadronic mass states determined by the light-front
Lorentz-invariant Hamiltonian equation for the relativistic bound-state system
$P_\mu P^\mu \vert  \psi(P) \rangle = M^2 \vert  \psi(P) \rangle$,
where the operator $P_\mu P^\mu$ is determined canonically from the QCD Lagrangian.
On AdS space the physical  states are
represented by normalizable modes $\Phi_P(x, z) = e^{-iP \cdot x} \Phi(z)$,
with plane waves along Minkowski coordinates $x^\mu$ and a profile function $\Phi(z)$
along the holographic coordinate $z$. The hadronic invariant mass
$P_\mu P^\mu = M^2$  is found by solving the eigenvalue problem for the
AdS wave equation. Each  light-front hadronic state $\vert \psi(P) \rangle$ is dual to a normalizable string mode $\Phi_P(x,z)$.
For fields near the AdS boundary the behavior of $\Phi(z)$
depends on the scaling dimension of corresponding interpolating operators.

We have shown recently a remarkable
connection between the description of hadronic modes in AdS space and
the Hamiltonian formulation of QCD in physical space-time quantized
on the light-front at equal light-front time  $\tau$.~\cite{deTeramond:2008ht}  Indeed, one may take  the LF bound state Hamiltonian equation of motion in QCD as a starting point to  derive  relativistic wave equations in terms of an invariant transverse variable $\zeta$, which measures the
separation of the quark and gluonic constituents within the hadron
at the same LF time. The result is a single-variable light-front relativistic
Schr\"odinger equation,  which is
equivalent to the equations of motion which describe the propagation of spin-$J$ modes in a fixed  gravitational background asymptotic to AdS space. Its eigenvalues give the hadronic spectrum and its eigenmodes represent the probability distribution of the hadronic constituents at a given scale.  Remarkably, the AdS equations correspond to the kinetic energy terms of  the partons inside a hadron, whereas the interaction terms build confinement and
correspond to the truncation of AdS space in an effective dual gravity  approximation.~\cite{deTeramond:2008ht}

Light-front holographic mapping  was originally obtained
by matching the expression for electromagnetic current matrix
elements in AdS space with the corresponding expression for the
current matrix element using light-front  theory in physical space
time.~\cite{Brodsky:2006uqa, Brodsky:2007hb} More recently we have shown that one
obtains the identical holographic mapping using the matrix elements
of the energy-momentum tensor.~\cite{Brodsky:2008pf}

\section{A Semiclassical Approximation to QCD\label{LFholog}}

We start with the QCD light-front Hamiltonian equation for a relativistic bound state
 $\vert \psi \rangle$ \begin{equation} \label{LFH}
P_\mu P^\mu \vert  \psi(P) \rangle =  M^2 \vert  \psi(P) \rangle,
\end{equation}
where $M^2$ is the invariant hadron mass
and  $P_\mu P^\mu  =  P^- P^+ -  \mbf{P}_\perp^2$.
We can compute $M^2$ from the hadronic matrix element
\begin{equation} \label{eq:Matrix}
\langle \psi(P') \vert P_\mu P^\mu \vert\psi(P) \rangle  =
M^2  \langle \psi(P' ) \vert\psi(P) \rangle,
\end{equation}
expanding the initial and final hadronic states in terms of their Fock basis of non interacting components: $\vert \psi \rangle = \sum_n \psi_n \vert n \rangle$.
The matrix element can then be
expressed as a sum of overlap integrals with diagonal elements for the non interacting terms
in the LF Hamiltonian. We find~\cite{deTeramond:2008ht}
\begin{equation}   \label{Mbj}
 M^2  =  \sum_n  \prod_{j=1}^{n-1} \int d x_j \, d^2 \mbf{b}_{\perp j} \,
\psi_n^*(x_j, \mbf{b}_{\perp j})
 \sum_q \left(\frac{ \mbf{- \nabla}_{ \mbf{b}_{\perp q}}^2 + m_q^2 }{x_q} \right)
 \psi_n(x_j, \mbf{b}_{\perp j})
  + ({\rm interactions}) ,
 \end{equation}
where the light-front wave functions $\psi$ depend only  on the $n \! - \! 1 $ independent relative partonic coordinates,
the longitudinal momentum fraction $x_i = k_i^+/P^+$, the transverse impact variables $\mbf{b}_{\perp i}$
(canonical conjugate to the transverse momentum $\mbf{k}_{\perp i}$) and $\lambda_i$, the
projection of the constituent's spin along the $z$ direction. Momentum conservation requires
$\sum_{i=1}^n x_i =1$ and $\sum_{i=1}^n \mbf{b}_{\perp i}=0$.
The normalization is defined by
\begin{equation}  \label{eq:LFWFbnorm}
\sum_n  \prod_{j=1}^{n-1} \int d x_j d^2 \mbf{b}_{\perp j}
\left\vert \psi_n(x_j, \mbf{b}_{\perp j})\right\vert^2 = 1.
\end{equation}

To simplify the discussion we will consider a two-parton  bound state. In the limit
$m_q \to 0$
\begin{equation}  \label{Mb}
M^2  =  \int_0^1 \! \frac{d x}{x(1-x)} \int  \! d^2 \mbf{b}_\perp  \,
  \psi^*(x, \mbf{b}_\perp)
  \left( - \mbf{\nabla}_{\mbf{b}_ \perp}^2 \right)
  \psi(x, \mbf{b}_\perp)   +  ({\rm interactions}).
 \end{equation}

To identify the key variable in (\ref{Mbj}) we notice that the functional dependence  for a given Fock state is given in terms of its off-mass shell energy
$M^2 \! - M_n^2$, where $M_n^2  = \left( \sum_{i=1}^n k_i^\mu\right)^2$. For $n=2$, $M_{n=2}^2 = \frac{\mbf{k}_\perp^2}{x(1-x)}$.
Similarly, in impact space the relevant variable for a two-parton state is  $\zeta^2= x(1-x)\mbf{b}_\perp^2$.
As a result, to first approximation  LF dynamics  depend only on the boost invariant variable
$M_n$ or $\zeta,$
and hadronic properties are encoded in the hadronic mode $\phi(\zeta)$
from the relation,
\begin{equation} \label{eq:psiphi}
\psi(x,\zeta, \varphi) = e^{i L \varphi} X(x) \frac{\phi(\zeta)}{\sqrt{2 \pi \zeta}} ,
\end{equation}
thus factoring out the angular dependence $\varphi$ and the longitudinal, $X(x)$, and transverse mode $\phi(\zeta)$.
We choose the normalization of the LF mode $\phi(z) = \langle \zeta\vert \psi\rangle$ as $\langle \phi \vert \phi \rangle =
\int d\zeta \, \vert \langle  \zeta \vert \phi \rangle \vert^2 = 1$.

We can write the Laplacian operator in (\ref{Mb}) in circular cylindrical coordinates
$(\zeta, \varphi)$ with $ \zeta = \sqrt{x(1-x)} \vert \mbf{b}_\perp \vert$:
$\nabla_\zeta^2 = \frac{1}{\zeta} \frac{d}{d\zeta} \left( \zeta \frac{d}{d\zeta} \right)
+ \frac{1}{\zeta^2} \frac{\partial^2}{\partial \varphi^2}$,
and factor out the angular dependence of the
modes in terms of the $SO(2)$ Casimir representation $L^2$ of orbital angular momentum in the
transverse plane.
Using  (\ref{eq:psiphi}) we find~\cite{deTeramond:2008ht}
\begin{equation} \label{eq:KV} \nonumber
M^2  =  \int \! d\zeta \, \phi^*(\zeta) \sqrt{\zeta}
\left( -\frac{d^2}{d\zeta^2} -\frac{1}{\zeta} \frac{d}{d\zeta}
+ \frac{L^2}{\zeta^2}\right)
\frac{\phi(\zeta)}{\sqrt{\zeta}}
+ \int \! d\zeta \, \phi^*(\zeta) U(\zeta) \phi(\zeta),
 \end{equation}
where $L = L^z$.  In writing the above equation we have summed up the complexity of the interaction terms in the QCD Lagrangian  in the addition of the effective
potential $U(\zeta)$, which is then modeled to enforce confinement at some IR scale.
The light-front eigenvalue equation $P_\mu P^\mu \vert \phi \rangle = M^2 \vert \phi \rangle$
is thus a light-front wave equation for $\phi$
\begin{equation} \label{LFWE}
\left(-\frac{d^2}{d\zeta^2}
- \frac{1 - 4L^2}{4\zeta^2} + U(\zeta) \right)
\phi(\zeta) = M^2 \phi(\zeta),
\end{equation}
an effective single-variable light-front Schr\"odinger equation which is
relativistic, covariant and analytically tractable.  Its eigenmodes $\phi(\zeta) = \langle \zeta \vert \phi \rangle$
determine the hadronic mass spectrum and represent the probability
amplitude to find $n$-partons at transverse impact separation $\zeta$,
the invariant separation between pointlike constituents within the hadron~\cite{Brodsky:2006uqa} at equal
light-front time. Extension of the results to arbitrary $n$ follows from the $x$-weighted definition of the
transverse impact variable of the $n-1$ spectator system:~\cite{Brodsky:2006uqa}
\begin{equation} \label{zeta}
\zeta = \sqrt{\frac{x}{1-x}} ~ \Big\vert \sum_{j=1}^{n-1} x_j \mbf{b}_{\perp j} \Big\vert ,
\end{equation}
where $x = x_n$ is the longitudinal
momentum fraction of the active quark. One can also
generalize the equations to allow for the kinetic energy of massive
quarks using  (\ref{Mbj}). In this case, however,
the longitudinal mode $X(x)$ does not decouple from the effective LF bound-state equations.

\section{Higher Spin Hadronic Modes in AdS Space}

The description of higher spin modes in AdS space is a notoriously difficult problem.~\cite{Fronsdal:1978vb, Fradkin:1986qy}
A spin-$J$ field in AdS$_{d+1}$ is represented by a rank $J$ tensor field $\Phi(x^A)_{M_1 \cdots M_J}$, which is totally symmetric in all its indices. Such a tensor contains also lower spins, which can be eliminated by imposing gauge conditions. The action for a spin-$J$ field in AdS$_{d+1}$ space time in presence of a dilaton background field $\varphi(z)$ is given by
\begin{eqnarray} \label{SJ} \nonumber
S = \half \int \! d^d x \, dz  \,\sqrt{g} \,e^{\varphi(z)}
  \Big( g^{N N'} g^{M_1 M'_1} \cdots g^{M_J M'_J} D_N \Phi_{M_1 \cdots M_J} D_{N'}  \Phi_{M'_1 \cdots M'_J}  \\
 - \mu^2  g^{M_1 M'_1} \cdots g^{M_J M'_J} \Phi_{M_1 \cdots M_J} \Phi_{M'_1 \cdots M'_J}  + \cdots \Big)  ,
\end{eqnarray}
where $D_M$ is the covariant derivative which includes parallel transport
\begin{equation} \label{Dco}
[D_N, D_K]  \Phi_{M_1 \cdots M_J} =  - R^L_{\, M_1 N K} \Phi_{L \cdots M_J} - \cdots  - R^L_{\, M_J N K} \Phi_{M_1 \cdots L},
\end{equation}
and the omitted terms refer to
terms with different contractions.  Conformal invariance in (\ref{SJ}) is broken by  $\varphi(z)$ which is a function of the holographic coordinate $z$ and vanishes
 in the conformal limit $z \to 0$.  The coordinates of AdS are the Minkowski coordinates $x^\mu$ and the holographic variable $z$ labeled $x^M = \left(x^\mu, z\right)$.

 A physical hadron has plane-wave solutions and polarization indices $\mu_i$, $i = 1 \cdots J$, along the 3 + 1 physical coordinates
 $\Phi_P(x,z)_{\mu_1 \cdots \mu_J} = e^{- i P \cdot x} \Phi(z)_{\mu_1 \cdots \mu_J}$,
 with four-momentum $P_\mu$ and  invariant hadronic mass  $P_\mu P^\mu \! = \! M^2$. All other components vanish identically:
 $\Phi_{z \mu_2 \cdots \mu_J} = \cdots = \Phi_{\mu_ 1 \mu_2 \cdots z} = 0$. One can then construct an effective action in terms
 of high spin modes $\Phi_J = \Phi_{\mu_1 \mu_2 \cdots \mu_J}$, with only the physical degrees of freedom.~\cite{Karch:2006pv, BDDE:2010xx} In this case the system of coupled differential equations which follow from (\ref{SJ}) reduce to a homogeneous equation in terms of the physical field $\Phi_J$.

We retain only physical modes $\Phi_{\mu_1 \mu_2 \cdots \mu_J}$, and start with the scalar wave equation which follows from the variation of (\ref{SJ}) for $J = 0$. This case is particularly simple as the covariant derivative of a scalar field is the usual derivative. We  obtain the eigenvalue equation
\begin{equation} \label{WeS}
\left[-\frac{ z^{d-1}}{e^{\varphi(z)}}   \partial_z \left(\frac{e^{\varphi(z)}}{z^{d-1}} \partial_z\right)
+ \left(\frac{\mu R}{z}\right)^2\right] \Phi = M^2 \Phi.
\end{equation}
A physical  spin-$J$ mode $\Phi_{\mu_1 \cdots \mu_J}$  with all  indices
along 3+1 is then constructed by shifting dimensions
$\Phi_J(z) = ( z/R)^{-J}  \Phi(z)$. Its   normalization is given by
\begin{equation}  \label{Phinorm}
R^{d - 1 - 2 J} \int_0^{\infty} \! \frac{dz}{z^{d -1 - 2 J}} \, e^{\varphi(z)} \Phi_J^2 (z) = 1.
\end{equation}
The shifted field $\Phi_{\mu_1 \mu_2 \cdots \mu_J}$ obeys the wave equation~\cite{deTeramond:2008ht, deTeramond:2010ge}
\begin{equation} \label{WeJ}
\left[-\frac{ z^{d-1 -2 J}}{e^{\varphi(z)}}   \partial_z \left(\frac{e^{\varphi(z)}}{z^{d-1 - 2 J}} \partial_z\right)
+ \left(\frac{\mu R}{z}\right)^2\right] \Phi_{\mu_1 \mu_2 \cdots \mu_J} = M^2 \Phi_{\mu_1 \mu_2 \cdots \mu_J},
\end{equation}
which follows from (\ref{WeS})
upon mass rescaling $(\mu R)^2 \to (\mu R)^2 - J(d-J)$  and $M^2 \to M^2 -  J z^{-1} \partial_z \varphi$.
For $J=1$ our results are identical with the wave equation for a massive AdS vector field in presence of a dilaton background.

\section{Light-Front Holographic Mapping and Hadronic Spectrum}

The structure of the QCD Hamiltonian equation  (\ref{LFH}) is similar to the structure of the AdS wave equation (\ref{WeJ}); they are both frame-independent and have identical eigenvalues $M^2$, the mass spectrum of the color-singlet states of QCD, a possible indication of a more profound connection between physical QCD and the physics of hadronic modes in AdS space. However, important differences are also apparent:  Eq. (\ref{LFH}) is a linear quantum-mechanical equation of states in Hilbert space, whereas Eq. (\ref{WeJ}) is a classical gravity equation; its solutions describe spin-$J$ modes propagating in a higher dimensional
warped space. Physical hadrons are composite and thus inexorably endowed of orbital angular momentum. Thus, the identification
of orbital angular momentum is of primary interest in finding a connection between both approaches.

As shown in the Sect. \ref{LFholog}, one can indeed systematically reduce  the LF  Hamiltonian eigenvalue Eq.  (\ref{LFH}) to an effective relativistic wave equation, analogous to the AdS equations, by observing that each $n$-particle Fock state has an essential dependence on the invariant mass of the system  and
thus, to a first approximation, LF dynamics depend only on $M_n^2$.
In  impact space the relevant variable is the boost invariant  variable $\zeta$ (\ref{zeta})
 which measures the separation of the  constituents and which also allows one to separate the dynamics
of quark and gluon binding from the kinematics of the constituent
internal angular momentum.

Upon the substitution $z \! \to\! \zeta$  and
$\phi_J(\zeta)   = \left(\zeta/R\right)^{-3/2 + J} e^{\varphi(z)/2} \, \Phi_J(\zeta)$,
in (\ref{WeJ}), we find for $d=4$ the QCD light-front wave equation (\ref{LFWE}) with effective potential~\cite{deTeramond:2010ge}
\begin{equation} \label{U}
U(\zeta) = \half \varphi''(z) +\frac{1}{4} \varphi'(z)^2  + \frac{2J - 3}{2 z} \varphi'(z) ,
\end{equation}
The fifth dimensional mass $\mu$ is not a free parameter but scales as $(\mu R)^2 = - (2-J)^2 + L^2$.

If $L^2 < 0$ the LF Hamiltonian  is unbounded from below
$\langle \phi \vert H_{LF} \vert \phi \rangle <0$  and the spectrum contains an
infinite number of negative values of $M^2 $, which can be arbitrarily large.
The critical value  $L=0$  corresponds to the lowest possible stable solution, the ground state of the light-front Hamiltonian.
For $J = 0$ the five dimensional mass $\mu$
 is related to the orbital  momentum of the hadronic bound state by
 $(\mu R)^2 = - 4 + L^2$ and thus  $(\mu R)^2 \ge - 4$. The quantum mechanical stability condition $L^2 \ge 0$ is thus equivalent to the
 Breitenlohner-Freedman stability bound in AdS.~\cite{Breitenlohner:1982jf}
The scaling dimensions are $2 + L$ independent of $J$ in agreement with the
twist-scaling dimension of a two-parton bound state in QCD.
It is important to notice that in the light-front the $SO(2)$ Casimir for orbital angular momentum $L^2$
is a kinematical quantity, in contrast with the usual $SO(3)$ Casimir $L(L+1)$ from non-relativistic physics which is
rotational, but not boost invariant.

We consider here the  positive-sign dilaton profile $\exp(+ \kappa^2 z^2)$ which confines the constituents  to distances
$\langle z \rangle \sim 1/\kappa$.~\cite {deTeramond:2009xk, Andreev:2006ct}
From (\ref{U}) we obtain  the effective potential~\cite{deTeramond:2009xk}
$U(\zeta) =   \kappa^4 \zeta^2 + 2 \kappa^2(L + S - 1)$,  where $J^z = L^z + S^z$, which  corresponds  to a transverse oscillator in the light-front.
Equation  (\ref{LFWE}) has eigenfunctions
\begin{equation} \label{phi}
\phi_{n, L}(\zeta) = \kappa^{1+L} \sqrt{\frac{2 n!}{(n\!+\!L\!)!}} \, \zeta^{1/2+L}
e^{- \kappa^2 \zeta^2/2} L^L_n(\kappa^2 \zeta^2) ,
\end{equation}
and eigenvalues
\begin{equation} \label{M2}
M_{n, L, S}^2 = 4 \kappa^2 \left(n + L + \frac{S}{2} \right).
\end{equation}

\begin{figure}[h]
\centering
\includegraphics[angle=0,width=4.6cm]{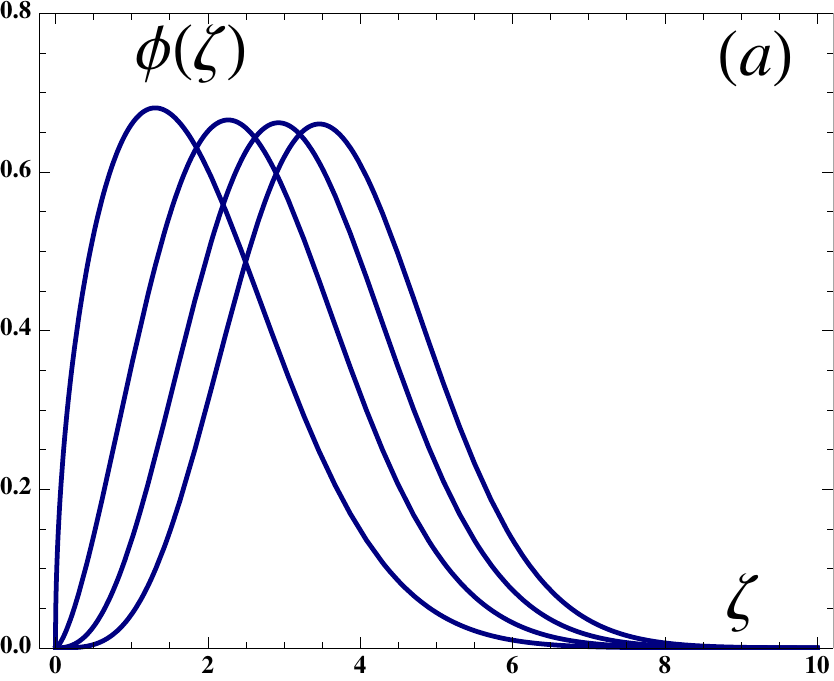} \hspace{20pt}
\includegraphics[angle=0,width=4.7cm]{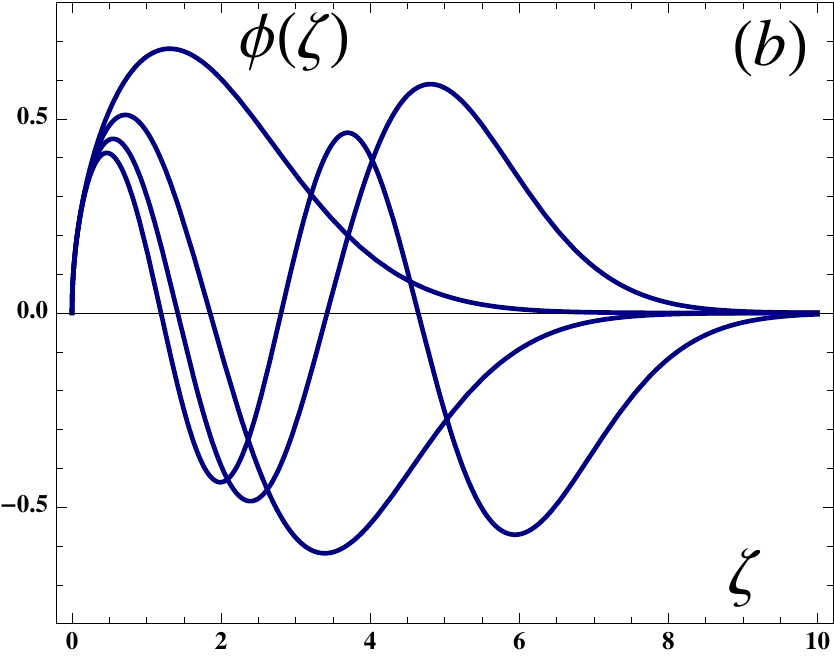}
\caption{Light-front wavefunctions $\phi_{n,L}(\zeta)$  is physical spacetime corresponding to a dilaton $\exp(\kappa^2 z^2)$: a) orbital modes ($n=0$) and b)
radial modes ($L=0$).}
\label{LFWFs}
\end{figure}

The meson spectrum  has a string-theory Regge form: the square of the masses are linear in both the internal orbital angular momentum $L$ and radial quantum number $n$, where $n$ counts the number of nodes  of the wavefunction in the radial variable $\zeta$. The spectrum also depends on the internal spin S.
The lowest possible solution for $n = L = S = 0$ has eigenvalue $M^2 = 0$.
This is a chiral symmetric bound state of two massless quarks with scaling dimension 2 and size
 $\langle \zeta^2 \rangle \sim 1/\kappa^2$, which we identify with the lowest state, the pion.
Thus one can compute the hadron spectrum by simply adding  $4 \kappa^2$ for a unit change in the radial quantum number, $4 \kappa^2$ for a change in one unit in the orbital quantum number and $2 \kappa^2$ for a change of one unit of spin to the ground state value of $M^2$. Remarkably, the same rule holds for baryons.~\cite{deTeramond:2009xk}  This is an important feature of light-front holography, which predicts the same multiplicity of states for mesons
and baryons  as it is observed experimentally.~\cite{Klempt:2007cp}
The LFWFs (\ref{phi}) for different orbital and radial excitations are depicted in Fig. \ref{LFWFs}.

\begin{figure}[h]
\begin{center}
\includegraphics[width=7.0cm]{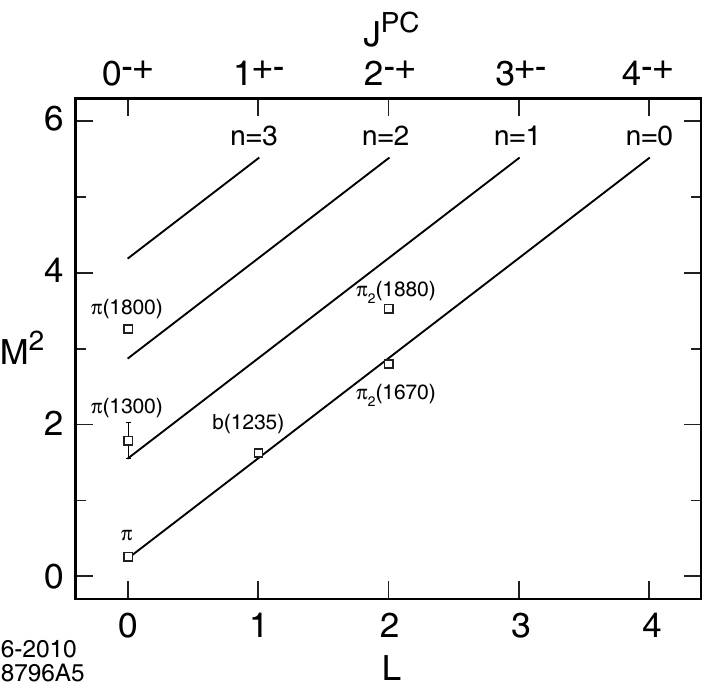}  \hspace{10pt}
\includegraphics[width=7.0cm]{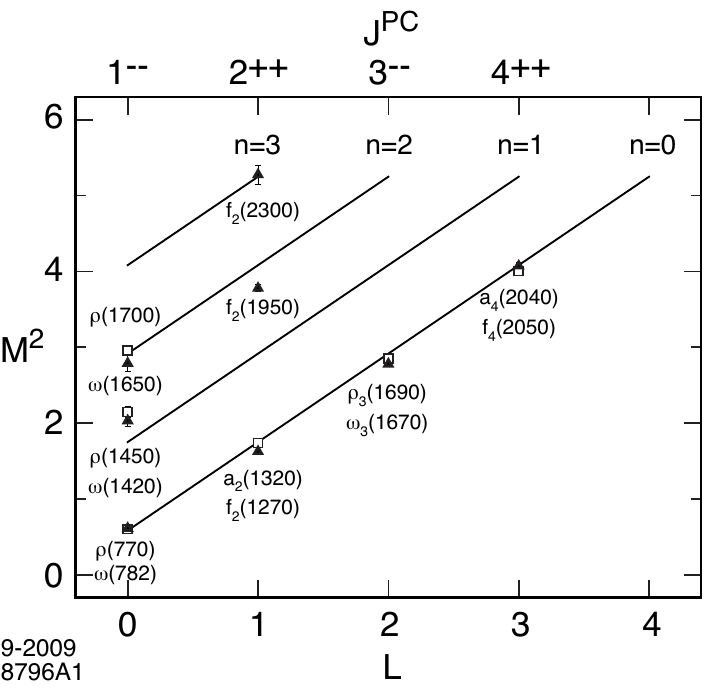}
 \caption{Parent and daughter Regge trajectories for (a) the $\pi$-meson family with
$\kappa= 0.6$ GeV; and (b) the  $I\!=\!1$ $\rho$-meson
 and $I\!=\!0$  $\omega$-meson families with $\kappa= 0.54$ GeV. Only confirmed PDG states~\cite{Amsler:2008xx} are shown.}
\label{pionspec}
\end{center}
\end{figure}

Individual hadron states are identified by their interpolating operators at $z\to 0.$ Pion interpolating operators are constructed by examining the behavior of
bilinear covariants $\bar \psi \Gamma \psi$ under charge conjugation and parity transformation.
Thus, for example, a pion interpolating operator $\bar q \gamma_5 q$ create a state with quantum numbers $J^{PC} = 0^{- +}$, and a vector meson
interpolating operator $\bar q \gamma_\mu q$ a state $1^{- -}$. Likewise the operator $\bar q \gamma_\mu \gamma_5 q$ creates a state with
$1^{++}$ quantum numbers,  the $a_1(1260)$ positive parity meson. If we include  orbital excitations the pion interpolating operator is
$\mathcal{O}_{2+L} = \bar q \gamma_5  D_{\{\ell_1} \cdots D_{\ell_m\}} q$. This is an operator  with total internal  orbital
momentum $L = \sum_{i=1}^m \ell_i$, twist $\tau = 2 + L$ and canonical dimension $\Delta = 3 + L$.  The scaling of  the AdS field $\Phi(z) \sim z^\tau$ at $z \to 0$  is precisely the scaling required to match the scaling dimension of the local meson interpolating operators.   The spectral predictions for  light meson and vector meson  states are compared with experimental data
in Fig. \ref{pionspec} for the positive sign dilaton model discussed here.

\section{Higher Fock Components in Light Front Holography}

The light front holographic variable $\zeta$ (\ref{zeta}) is particularly useful in describing a multiple parton state, as it incorporates a cluster decomposition: one particle (the active quark) {\it vs.} the rest (the spectator system). Thus, for example, for a baryon the LF cluster decomposition is equivalent to a quark-diquark system, and this may explain why LF holography is successful in  predicting the same multiplicity of meson and baryon states.~\cite{deTeramond:2009xk}

The LF Hamiltonian eigenvalue equation (\ref{LFH}) is a matrix in Fock space. Writing   $P_\mu P^\mu \equiv H_{LF}$ as a sum of  terms representing the  kinetic energy of the partons $H^0_{LF}$ plus an interaction
 potential $V$, $H_{LF} = H^0_{LF} + V$, we find upon expanding in Fock eigenstates of $H^0_{LF}$, $\vert \psi \rangle = \sum_n \psi_n \vert n \rangle$,
\begin{equation}
\left(M^2 - \sum_{i=1}^n \frac{\mbf{k}^2_{ \perp i} + m^2}{ x_i} \right) \psi _n  =
  \sum_m \langle  n  \vert V \vert m \rangle \psi _m ,
 \end{equation}
which represents an infinite
number of coupled integral equations.  In AdS/QCD the only interaction  is the confinement potential. The resulting potential in quantum field theory is the four-point effective interaction
$H_I ={\overline \psi} \psi ~ V \!\left(\zeta^2\right) {\overline \psi }\psi$, 
which leads to $q q \to q q$ , $q \bar q \to q \bar q$, $ q\to q  q \bar q$ and  $\bar  q\to \bar  q  q \bar q$, thus creating
states with extra quark-antiquark pairs. In this approximation there is no mixing with  the $q \bar q g$ Fock states  from the interaction term $g_s {\overline \psi} \gamma \cdot A \psi$ in QCD. Since models based on AdS/QCD are particularly successful in the description of exclusive processes,~\cite{Brodsky:2010cq}
this may explain the dominance of quark interchange~\cite{Gunion:1972qi}
 over quark annihilation or gluon exchange contributions in large angle elastic scattering.~\cite{Baller:1988tj}

 To show the relevance of higher Fock states we discuss in the next section a simple semi-phenomenological model where we include the first two components in a Fock expansion of the pion wave function
$\vert \pi \rangle  = \psi_{q \bar q /\pi} \vert q \bar q  \rangle_{\tau=2}
+  \psi_{q \bar q q \bar q} \vert q \bar q  q \bar q  \rangle_{\tau=4} + \cdots$ ,
where the $J^{PC} = 0^{- +}$ twist-two and twist-4 states $\vert q \bar q \rangle$  and  $\vert q \bar q q \bar q  \rangle$ are created by the interpolating operators
$\bar q \gamma_5 q$ and $ \bar q \gamma_5 q \bar q q$ respectively.

\section{Space- and Time-Like Structure of the Pion Form Factor}

\begin{figure}[h]
\begin{center} \label{pionFF}
\includegraphics[width=6.45cm]{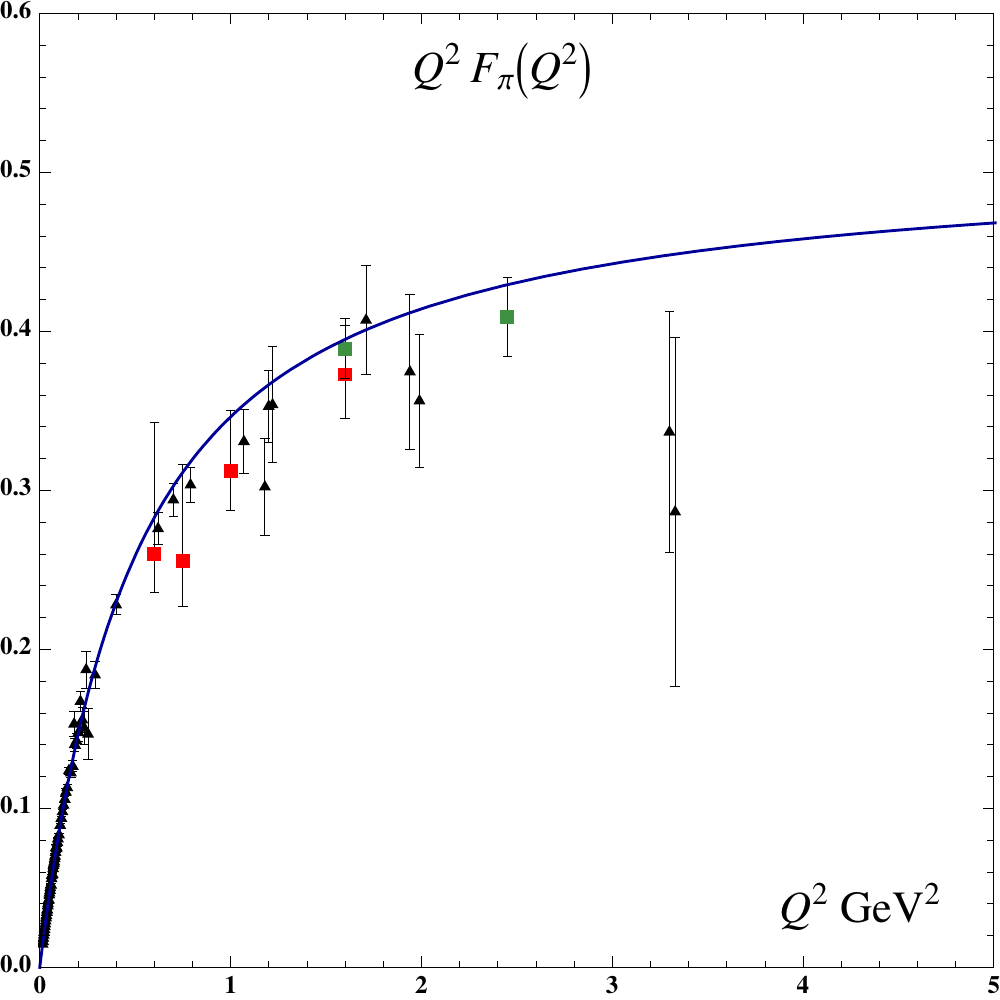} \hspace{8pt}
\includegraphics[width=7.10cm]{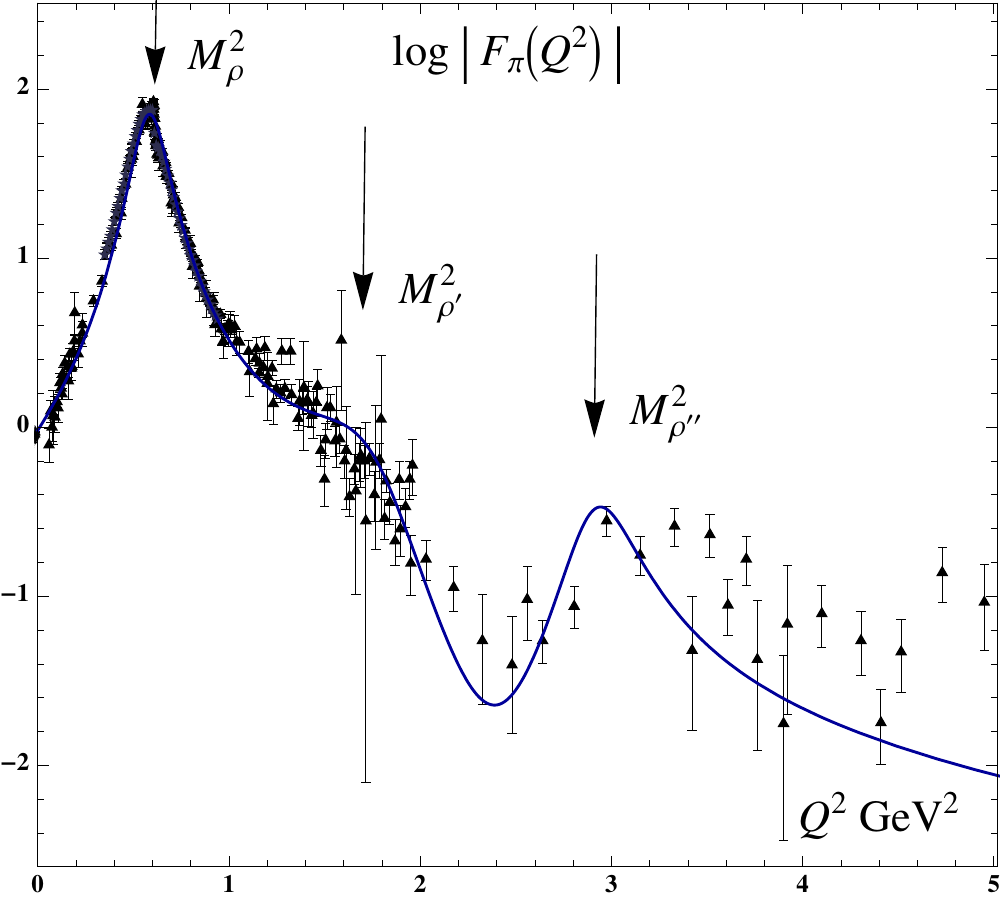}
\caption{Structure of the space- and time-like pion form factor in light-front holography for a truncation of the pion wave function up to twist four.
Triangles are the data compilation  from Baldini  {\it et al.},~\cite{Baldini:1998qn} red squares are JLAB 1~\cite{Tadevosyan:2007yd} and green squares are JLAB 2.~\cite{Horn:2006tm}}
\end{center}
\end{figure}

In the soft wall model  the electromagnetic probe propagates in modified  AdS metrics. As a result the current is dual to a dressed current, {\it i.e.}, a hadronic electromagnetic current including virtual $\bar q q$ pairs and thus confined.  In this case, the bulk-to-boundary propagator  $J(Q,z)$ has the integral representation~\cite{Grigoryan:2007my}
\begin{equation} \label{Jkappa}
J(Q,z) = \kappa^2 z^2 \int_0^1 \! \frac{dx}{(1-x)^2} \, x^{\frac{Q^2}{4 \kappa^2}}
e^{-\kappa^2 z^2 x/(1-x)}.
\end{equation}
The form factor corresponding to (\ref{Jkappa}) for a state with twist $\tau = N$, is expressed  as an $N - 1$ product of poles, corresponding to the first $N-1$ states along the vector meson radial trajectory~\cite{Brodsky:2007hb}
\begin{equation} \label{FF}
F(Q^2) =  \frac{1}{\Big(1 + \frac{Q^2}{\mathcal{M}^2_\rho} \Big)
 \Big(1 + \frac{Q^2}{\mathcal{M}^2_{\rho'}}  \Big)  \cdots
       \Big(1  + \frac{Q^2}{\mathcal{M}^2_{\rho^{N-2}}}  \Big)}.
\end{equation}
For a pion, for example, the lowest Fock state  -- the valence state -- is a twist 2 state, and thus the form factor is the well known  monopole form.~\cite{Brodsky:2007hb}  Since the charge form factor is a diagonal operator, the final expression for the form factor corresponding to the truncation up to twist four is the sum of two terms, a monopole and a three-pole term.
In the strongly coupled semiclassical gauge/gravity limit hadrons have zero widths and are stable. One can nonetheless modify the formula (\ref{FF}) to introduce a finite width:
$q^2 \to q^2 + 2 i \kappa \Gamma$.  We choose the values $\Gamma_\rho =  130$ MeV,   $\Gamma_\rho =  400$ MeV and  $\Gamma_\rho =  300$ MeV.  The results for the pion form factor with higher Fock states (twist two and four) are shown in Fig. (\ref{pionFF}). The results correspond to $P_{q \bar q q \bar q}$ = 13 \%, the admixture of the
$\vert q \bar q q \bar q  \rangle$ state. The value of $P_{q \bar q q \bar q}$ (and the widths) are input in the model. The value of $\kappa$ is determined from the $\rho$ mass and the masses of the radial excitations follow from (\ref{M2}). The time-like structure of the pion form factor displays a rich pole structure with constructive and destructive interferences.

Conserved currents  correspond to five dimensional massless fields in AdS according to the relation 
$(\mu R)^2 = (\Delta - p) (\Delta + p -  4)$  for a $p$ form in $d=4$. In the usual AdS/QCD framework~\cite{Erlich:2005qh, DaRold:2005zs} this  corresponds to $\Delta = 3$ or 1, the canonical dimensions of
an EM current and field strength respectively.  Normally one uses a hadronic  interpolating operator  with minimum twist $\tau$ to identify a hadron  to predict the power-law fall-off behavior of its form factors and other hard 
scattering amplitudes;~\cite{Polchinski:2001tt} {\it e.g.},  for a two-parton bound state $\tau = 2$.   However, in the case of a current, one needs to  use  an effective field operator  with dimension $\Delta =3.$ The apparent inconsistency between twist and dimension is removed by noticing that in the light-front one chooses to calculate the  matrix element of the twist-3 plus  component of the current  $J^+$,~\cite{Brodsky:2006uqa, Brodsky:2007hb} in order to avoid coupling to Fock states with different numbers of constituents.

\section{Conclusions}

Light-front  holography provides a direct correspondence between an effective gravity theory defined in a fifth-dimensional warped space and a  semiclassical approximation to strongly coupled QCD quantized on the light-front. This duality leads to a remarkable Lorentz-invariant relativistic Schr\"odinger-like equation~\cite{deTeramond:2008ht} which
provides a successful prediction for the light-quark meson and baryon spectra as
a function of hadron spin, quark angular momentum, and radial quantum numbers. It also predicts the same multiplicity of states for mesons
and baryons, which is observed experimentally.
We originally derived this correspondence using the identity between electromagnetic and gravitational form factors computed in AdS and light-front theory.~\cite{Brodsky:2006uqa,Brodsky:2007hb,Brodsky:2008pf}  The results for hadronic form factors are also successful, and the predicted power law fall-off agrees with dimensional counting rules as required by conformal invariance at small $z$.~\cite{Brodsky:2007hb, Brodsky:2008pg} As in the Schr\"odinger equation, the semiclassical approximation to light-front QCD described in this paper does not account for particle creation and absorption; it is thus expected to break down at short distances
where hard gluon exchange and quantum corrections become important.
However, one can systematically improve the semiclassical approximation, for example by introducing nonzero quark masses and short-range Coulomb
corrections.~\cite{Branz:2010ub, Arriola:2010up}  We have discussed the relevance of higher Fock-states for describing the detailed structure of  form factors. A simple model including
twist-two and twist-four Fock components for the pion wavefunction describes remarkable well the pole structure and the effects of constructive and destructive interferences in the time-like region.

The hadron eigenstate generally has components with different orbital angular momentum.  For example, the proton eigenstate in light-front holography with massless quarks has $L=0$ and $L=1$ light-front Fock components with equal probability -- a novel manifestation of chiral invariance.~\cite{Brodsky:2010px}
Light-front holographic mapping of  effective classical gravity in AdS space, modified by the positive-sign dilaton background,  predicts the form of a non-perturbative effective coupling $\alpha_s(Q)$ and its $\beta$-function.~\cite{Brodsky:2010ur} The AdS running coupling is in very good agreement with the effective
coupling  extracted from  the Bjorken sum rule.~\cite{Deur:2008rf} The holographic $\beta$-function displays a transition from  nonperturbative to perturbative  regimes  at a momentum scale $Q \sim 1$ GeV.

\vspace{20pt}

\noindent{\bf \large Acknowledgements}

\vspace{10pt}

We thank  Alexander Deur, Josh Erlich and Hans Guenter-Dosch for collaborations. GdT thanks the members of the High Energy Physics Group at Imperial College in London for their hospitality. Invited talk presented by GdT at Light Cone 2010:  Relativistic Hadronic and Particle Physics, 14-18 June 2010, Valencia, Spain.  We are grateful to the organizers for  their outstanding hospitality.  This work was supported by Fondo de Incentivos CONICIT/MICIT, Costa Rica  and by the Department of Energy  contract DE--AC02--76SF00515. SLAC-PUB-14259.

\end{document}